# The omnidirectional receiver for UWOC systems based on the diffractive deep neural network

**JIANMIN XIONG,**[1,3] **JINGXUAN CHENG,**[1,3] **HUAN DENG,**[1,3] **YAN HUA,**[1,3] **YUFAN ZHANG,**[1,3] **ZIHAO DU,**[1,3] **LYUFANG ZHAO,**[1,3] **ZEJUN ZHANG,**[1,3] **AND JING XU**[1,2,3,*]

[1] *Optical Communications Laboratory, Ocean College, Zhejiang University, Zheda Road 1, Zhoushan, Zhejiang, 316021, China*
[2] *Hainan Institute of Zhejiang University, Sanya, China*
[3] *ZTT-Ocean College Joint Research Center for Marine Optoelectronic Technology, Ocean College, Zhejiang University, Zheda Road 1, Zhoushan, Zhejiang 316021, China*
*\* jxu-optics@zju.edu.cn*

**Abstract:** The link alignment requirement in underwater wireless optical communication (UWOC) systems is a knotty problem. The diffractive deep neural network ($D^2NN$) has shown great potential in accomplishing tasks all optically these years. In this paper, an omnidirectional receiver based on 7-layer $D^2NN$ is first proposed to alleviate the link alignment difficulties in UWOC systems. In addition, the vectorial diffraction theory is introduced into the training of the $D^2NN$ to obtain more accurate diffraction calculations compared with the prevalently adopted scalar diffraction theory. Simulation results verify the validity of the vectorial diffraction theory and demonstrate that the presented method can focus incident light waves with tilt angles from 0 to 89 degrees in a 6.25% area of the detection plane with an average focusing efficiency of 90.96%, proving the feasibility of omnidirectional focusing at the receiver end. Extra simulations further reveal that more layers do not lead to a sustained performance improvement but rather reach a bottleneck, and the $D^2NN$ can achieve omnidirectional focusing with a certain range of focusing region size. With more effort, the proposed receiver design, which can be highly integrated with detectors, holds promise to realize both omnidirectional and reliable link establishment in UWOC systems in the future.

## 1. Introduction

Due to the intricate and volatile underwater environment, the optical signals are disturbed by the absorption, scattering, and turbulence effects of the water during the underwater propagation [1,2]. Besides, the positions of the transmitter and receiver are continuously altered by water fluctuations. Therefore, it is challenging to maintain a stable establishment of the communication link in underwater wireless optical communication (UWOC) systems. Moreover, long-range UWOC systems have high restrictions on the beam divergence angle, which also adds to the difficulty of link alignment in practical scenarios [3]. Many related works have been proposed for now [4-14], which can be divided into two parts including the transmitter and receiver.

In the design of omnidirectional transmitters, both the light-emitting diodes (LEDs) and laser diodes (LDs) were adopted to achieve the rough and precise node positioning, respectively [4]. A prismatic array of three uniformly spaced high-power LED modules was made to realize quasi-omnidirectional radiation [5]. An optical diffuser and LD were combined to obtain a larger coverage area [6]. The photoluminescence of perovskite quantum dots was introduced to the design of a quasi-omnidirectional transmitter [7]. By applying the freeform lens to a LED array, a transmitter with a divergence angle of 150 degrees was proposed [8]. The omnidirectional design at the transmitter end can partially reduce the alignment requirements while leading to a lower energy efficiency, which needs a trade-off in different situations and is not the main focus of this paper.

On the receiver side, the use of a detector array is an effective way to alleviate the link alignment difficulties in UWOC systems. For example, high-sensitive multi-pixel photon counters were used to extend the misalignment tolerance [9]. Besides, the series-connected solar arrays were adopted for a large detection area and a high data transmission rate [10]. The plastic scintillating fibers were introduced into the large-area photoreceiver design [11-13]. Machine vision-based tracking was also proposed to achieve dynamic link alignment [14]. In these above-presented methods, some are only applicable in a certain angular range and some cannot be highly integrated with detectors, leading to the urgency of achieving both omnidirectional and reliable optical signal detection at the receiver end.

With the boom in deep learning and optics these years, optical computing is considered a competitive candidate for the next generation of computing mechanisms [15,16]. And the optical neural network, which was first explicitly proposed in the 1980s [17], has regained widespread attention from the scientific community. The optical implementation of neural networks has always been a compelling topic, attributed to the tremendous potential offered by the unique properties of optical waves including high speed, high bandwidth, robust parallelism, and low energy consumption, which can alleviate the inherent deficiencies of conventional Von Neumann computing architecture, such as memory wall and power wall [15,16].

Among various fulfillments of optical neural networks, the diffractive deep neural network ($D^2NN$) [18] is a remarkable one that performs complex pattern recognition tasks all optically and presents the synergy between deep learning algorithm acceleration and optical inverse design [19]. The primary function of the $D^2NN$ is to realize pixel-wise phase modulation of the incident light through successive layers. To further enhance the performance of the $D^2NN$, different operations have been adopted, for instance, differential detection [20], spatial transformation [21], and optical nonlinear material attachment [22]. These reported methods are effective in terms of performance improvement compared with the original $D^2NN$, but few studies notice that the scalar diffraction theory is theoretically unsuitable in the diffraction calculation of the $D^2NN$. Hence, the vectorial diffraction theory is introduced into the training of the $D^2NN$.

In the deployment of the $D^2NN$, the non-normal incidence of light is generally regarded as one of some undesired interferences that studies have been presented to mitigate [23,24]. Instead, in this paper, the non-normal incident light is the positive object to deal with. The $D^2NN$ is introduced into the receiver design to fully exploit the signal processing in the optical domain for both omnidirectional focusing and reliable link establishment which can be highly integrated with detectors in UWOC systems.

The rest of the paper is organized as follows. Section 2 introduces the equivalent treatment to non-normal incident illumination, the vectorial diffraction theory, and the diffraction calculation in the $D^2NN$. In section 3, the simulation setup is described in detail. Section 4 illustrates the simulation results, and section 5 includes some discussions. Eventually, section 6 concludes the paper.

## 2. Methods

*2.1 Equivalent treatment to non-normal incident illumination*

A light wave traveling a certain distance underwater is assumed to be a uniform plane wave at the receiver end in this paper. The position of the optical spot on the output plane generated by a focusing lens is sensitive to the direction of the incident light, as shown in Fig. 1(a). Consequently, stable omnidirectional focusing cannot be achieved by a focusing lens. In this work, the $D^2NN$ is introduced to handle the incident light with various tilt angles so that most of the output intensity can always be confined to a small fixed central region. For computational convenience, the non-normal incident light with different tilt angles is treated as plane waves with different phase modulation on the input plane before passing through multiple diffractive layers, as illustrated in Fig. 1(b).

With the center of the first diffractive layer as the coordinate origin, the zero-phase plane of the incident light with a tilt angle of $\alpha$ intersects with the input plane of the first diffractive layer. The distance $d$ from each point $(x, y)$ on the input plane to the zero-phase plane along the direction of the incident light determines the phase value of each point, as shown in Fig. 1(c). And the distance $d$ is determined by the angle $\beta$, whose value is equal to $\alpha$ in geometry, and the distance $l$ from the point $(x, y)$ to the intersection line (zero-phase line), as illustrated in Fig. 1(d). By solving the geometric problem, the equivalent phase modulation distribution $U$ of non-normal incident illumination can be written as

$$U = \exp(jkd)\sqrt{\cos\alpha}, \qquad (1)$$
$$d = -\sqrt{x^2 + y^2}\sin(\theta - \varphi)\sin\beta,$$

where $k = 2\pi/\lambda$, $x, y \in [-W/2, W/2]$, $\alpha = [0, \pi/2)$, $\theta \in [0, 2\pi)$, $\varphi \in [0, 2\pi)$, $\beta = \alpha$, $k$ is the wave number, $\lambda$ is the light wavelength, $W$ is the side length of the diffractive layer, $\theta$ is the rotation angle of the zero-phase line around the origin, and $\varphi$ is the angle of vector $(x, y)$ and x-coordinate. To normalize the input intensity based on the normal incident light and consider the finite power transmission carried by non-normal illumination, a cosine factor associated with the angle $\alpha$ is multiplied by the exponential term of $U$ in Eq. (1). The sign of the distance $d$ is determined by the relative direction between the zero-phase line and the point $(x, y)$, as shown in Fig. 1(d).

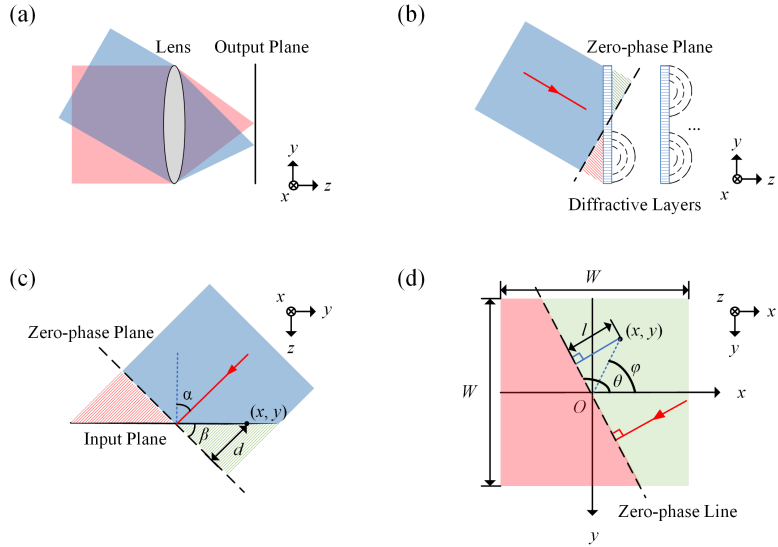

Fig. 1. Equivalent treatment to non-normal incident light. (a) Position variation of the light spot generated by a focusing lens. (b) The equivalent treatment to non-normal incident illumination before passing through multiple diffractive layers. (c)(d) Derivation of the equivalent phase modulation distribution of the non-normal incident light on the input plane of the first diffractive layer. The distance $d$ in the green region is negative and that in the red region is positive.

## 2.2 Vectorial diffraction theory

The compact interconnections in the D²NN are constructed based on the scalar diffraction theory, which greatly releases the burden of diffraction computation under the assumption that the aperture size is much larger than the light wavelength and the light travels a long distance before detection [25]. Although many works have presented effective improvements to enhance the performance of the D²NN [20-24], few studies notice a theoretical flaw in the diffraction calculation of the D²NN.

The implementation of full interconnections between adjacent diffractive layers requires a smooth diffraction output without bright alternating streaks, as shown in Fig. 2(a)(b), which imply that the aperture size of each point on the diffractive layer should be smaller than the light wavelength. It keeps consistent with the design of the original D²NN whose aperture size is 400 μm at the illumination wavelength of 0.75 mm [18]. With the requirement of diffraction aperture size, it is not appropriate to adopt the scalar diffraction theory to the diffraction calculation in the D²NN, because the coupling among electromagnetic wave components cannot be neglected when the light passes through an aperture whose size is smaller than its wavelength [26-28]. Hence, it is necessary to introduce the vectorial diffraction theory into the D²NN.

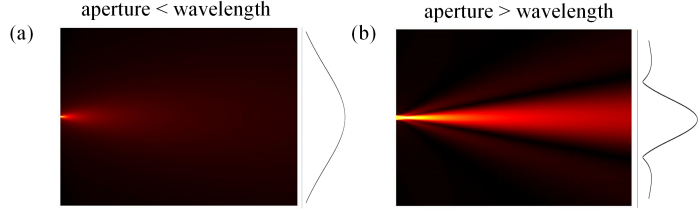

Fig. 2. Diffraction output of different aperture sizes at the illumination wavelength of 0.45 μm. (a) The aperture size of 0.2 μm. (b) The aperture size of 1 μm.

Although a research has proposed a similar opinion, the proposed construction itself was still based on the scalar diffraction theory [29]. And a study claimed that the scalar diffraction theory is accurate enough without needing the vectorial diffraction theory [30]. In this work, the vectorial diffraction theory is employed for the training of the D²NN and the simulation comparison between the two diffraction theories is performed to verify the correctness of the vectorial diffraction theory.

Similar to the derivation of the scalar diffraction theory, which is named the scalar Rayleigh-Sommerfeld diffraction integral (s-RSDI), the solution of the vectorial diffraction theory $\boldsymbol{E}(x,y,z) = \boldsymbol{i}E_x(x,y,z) + \boldsymbol{j}E_y(x,y,z) + \boldsymbol{k}E_z(x,y,z)$, which satisfies a specific input boundary condition $\boldsymbol{E}(\xi,\eta,0) = \boldsymbol{i}E_x(\xi,\eta,0) + \boldsymbol{j}E_y(\xi,\eta,0)$, can be obtained by solving the vectorial Helmholtz equation with the help of the Green theorem and some vector identities, which is called the vectorial Rayleigh-Sommerfeld diffraction integral (v-RSDI). Furthermore, an explicit form of the v-RSDI can be obtained under the condition that the divergence of the output electric field is equal to zero [26]:

$$E_x(x,y,z) = -\frac{1}{2\pi} \iint_\Sigma E_x(\xi,\eta,0) \frac{\partial}{\partial z}\left(\frac{\exp(jkR)}{R}\right) d\xi d\eta,$$

$$E_y(x,y,z) = -\frac{1}{2\pi} \iint_\Sigma E_y(\xi,\eta,0) \frac{\partial}{\partial z}\left(\frac{\exp(jkR)}{R}\right) d\xi d\eta, \qquad (2)$$

$$E_z(x,y,z) = \frac{1}{2\pi} \iint_\Sigma \left( E_x(\xi,\eta,0) \frac{\partial}{\partial x}\left(\frac{\exp(jkR)}{R}\right) + E_y(\xi,\eta,0) \frac{\partial}{\partial y}\left(\frac{\exp(jkR)}{R}\right) \right) d\xi d\eta,$$

where $k = 2\pi/\lambda$, $R = \sqrt{(x-\xi)^2 + (y-\eta)^2 + z^2}$, $d\xi$ and $d\eta$ are sampling intervals in the aperture plane $\Sigma$, $\boldsymbol{E}(\xi,\eta,0)$ and $\boldsymbol{E}(x,y,z)$ are the input and output electric fields passing through the aperture and propagating a distance of $z$, respectively. From Eq. (2), the v-RSDI is sensitive to the polarization state of the incident light, and the main difference between the s-RSDI and the v-RSDI is the output electric field component in the z-direction.

However, it is still challenging to directly calculate integrals of Eq. (2) by a computer, so proper simplification is necessary. Since the aperture diameter is smaller than the light wavelength, the complex amplitude reaching the aperture plane can be regarded as uniform. Therefore, the main task is how to deal with the simplification of $R$. On the one hand, the computation error is small in the near-axis region but will rise in the off-axis region if the calculation result on the whole aperture plane is directly replaced with that of the center point on the aperture plane, which is like the original D²NN. On the other hand, the calculation with a small sampling interval in the aperture plane will increase the calculation burden. Although many studies have been put forward to simplify the calculation of Eq. (2) [27-28], these methods are still quite intricate owing to their high computing complexity.

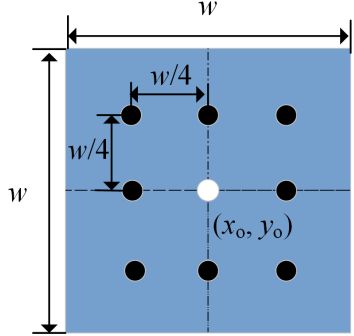

Fig. 3. The nine-point discrete sampling of the aperture plane. The center of the aperture plane is $(x_o, y_o)$. The side length of the aperture is $w$. The interval among sampling points is $w/4$.

To both maintain an accurate diffraction calculation and reduce the computation burden of Eq. (2), the method of nine-point discrete sampling is proposed by multiple simulations. In the diffraction calculation, symmetric nine points spaced at one-fourth of the aperture side length are chosen to represent the whole aperture plane, as shown in Fig. 3. For convenience, under the illumination of an x-polarized normal incident plane wave $\boldsymbol{E}(\xi,\eta,0)=\boldsymbol{i}$ with the wavelength $\lambda$, Eq. (2) can be rewritten as

$$E_x(x,y,z) = -\frac{w^2}{32\pi}\sum_{i=-1}^{1}\sum_{j=-1}^{1}\frac{z}{r}\left(jk-\frac{1}{r}\right)\frac{\exp(jkr)}{r},$$

$$E_z(x,y,z) = \frac{w^2}{32\pi}\sum_{i=-1}^{1}\sum_{j=-1}^{1}\frac{x-x_o-i\left(\frac{w}{4}\right)}{r}\left(jk-\frac{1}{r}\right)\frac{\exp(jkr)}{r}, \quad (3)$$

$$r = \sqrt{\left(x-x_o-i\left(\frac{w}{4}\right)\right)^2 + \left(y-y_o-j\left(\frac{w}{4}\right)\right)^2 + z^2}.$$

### 2.3 Diffractive deep neural network

The overall structure of the D²NN is illustrated in Fig. 4, and tasks are accomplished by optimizing the phase modulation parameters on these diffractive layers. In this work, various input plane waves with different tilt angles pass through multiple diffractive layers and finally, the intensity on the output plane is limited in a small fixed central region named the focusing region. Meanwhile, each plane is sampled as a grid region with a size of $N \times N$. The complex amplitude of each point on the latter layer can be regarded as the output superposition of each point on the former layer based on Eq. (3), which can be written as

$$U_{x,z}^l(x_m, y_n, z_l) = \sum_{i=1}^{N}\sum_{j=1}^{N}U_{x,z}^{l-1}(x_i, y_j, z_{l-1})g_{x,z}(x_m-x_i, y_n-y_j, z_l-z_{l-1}), \quad (4)$$

$$g_x(x,y,z) = E_x(x,y,z), \quad g_z(x,y,z) = E_z(x,y,z), \quad x_o = y_o = 0,$$

where $U^l_{x,z}$ and $U^{l-1}_{x,z}$ are complex amplitude values on the former and latter planes, respectively, in which $l$ is the current layer number and $x$, $z$ denotes different components. The points $(x_i, y_j)$ and $(x_m, y_n)$ are sampled on the former and latter planes. And $i, j, m, n = 1, 2, …, N$. The $z_l − z_{l-1}$ denotes the distance between two adjacent layers. In addition, Eq. (4) is a standard two-dimension discrete linear convolution between $U^{l-1}$ and $g$. Thus, a fast Fourier transform-based direct integration (FFT-DI) method is selected to get the numerical result of Eq. (4) fast and accurately, which is given as [31]:

$$O = \text{IFFT2}\{\text{FFT2}\{U\} * \text{FFT2}\{G\}\}, \tag{5}$$

where FFT2{·} and IFFT2{·} denote the two-dimension fast Fourier transform and inverse two-dimension fast Fourier transform, respectively. The matrix $U$ contains $U^{l-1}$ with zero padding to a size of $2N\text{-}1 \times 2N\text{-}1$, and $G$ is the matrix of $g$ with a size of $2N\text{-}1 \times 2N\text{-}1$. The main reason for zero padding is that the calculation in Eq. (5) is equal to the result of two-dimension circular convolution, but not two-dimension discrete linear convolution. The symbol * denotes the Hadamard product. Ultimately, the downright region of $O$ with a size of $N \times N$ is the target output. More details about the above calculation are well described in [31].

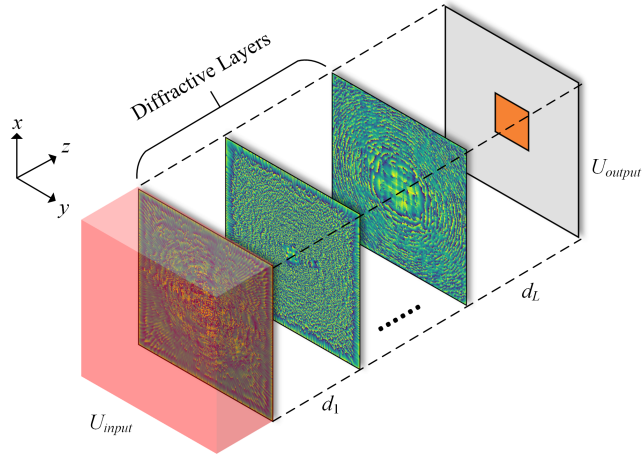

Fig. 4. The structure of the D²NN. The $U_{input}$ denotes the input optical complex amplitude, which is a uniform plane wave with a certain tilt angle in this study. The $U_{output}$ denotes the optical complex amplitude that finally reaches the output plane. The elements in $\{d_1, …, d_L\}$ denote the distances between every two adjacent layers, and the subscript $L$ denotes the number of diffractive layers.

## 3. Simulation setup

### 3.1 Structure design of the D²NN

In this paper, a 7-layer phase-only modulation D²NN is constructed to achieve omnidirectional focusing at the receiver end of UWOC systems. Ultimately, the output intensity is finally restricted in the small fixed central region, as illustrated in Fig. 4. The structural parameters of the proposed D²NN are listed in detail and shown in Table 1.

It is necessary to explain some structural parameter settings. The tilt angle $α$ is defined as the angle between the wave vector of illumination light and the z-coordinate, which is set from 0 to 89 degrees with an interval of 1 degree. In addition, for each equivalent phase modulation distribution on the input plane at a certain tilt angle shown in Fig. 1(d), the whole distribution is sampled whenever the rotation angle $θ$ varies in the range of [0º, 360º) with an interval of 1 degree to augment the dataset and increase the robustness of omnidirectional focusing. The interlayer distance is set to 40 times the light wavelength which can readily reach full interconnections among adjacent layers. As for the diffraction aperture size, more simulations are required to select a better one at the illumination wavelength of 0.45 μm.

Table 1. Structure parameters of the D²NN

| Parameter | Value |
| --- | --- |
| Illumination Wavelength ($\lambda$) | 0.45 μm |
| Diffraction Unit (Aperture) Size ($w$) | 0.2 μm |
| Layer Distance ($d_1, …, d_L$) | 18 μm |
| Number of Diffractive Layers ($L$) | 7 |
| Unit Number on each layer ($N \times N$) | 200×200 |
| Unit Number of Focusing Region ($S \times S$) | 50×50 |
| Range of Tilt Angle ($\alpha$) | [0°, 89°] |

As mentioned previously, an aperture smaller than the light wavelength is needed in the D²NN. Therefore, multiple diffraction outputs of apertures with different sizes are obtained by the finite difference time domain (FDTD) method. The magnitude of two-dimension diffraction outputs along the x-coordinate is shown in Fig. 5.

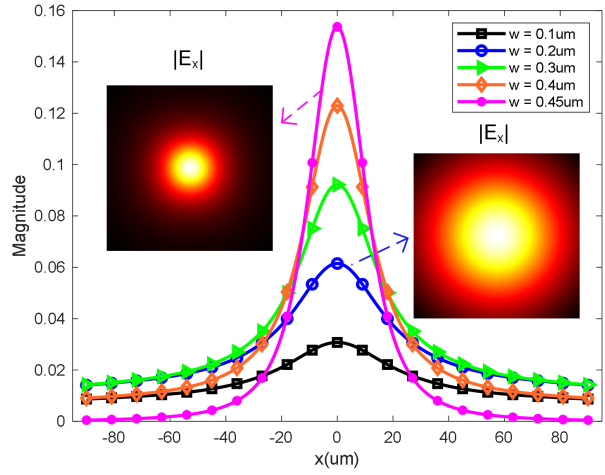

Fig. 5. Diffraction outputs of different aperture sizes with the illumination wavelength of 0.45 μm generated by the FDTD method. The two subgraphs are normalized two-dimension magnitudes of the component $|E_x|$ with an aperture size of 0.2 μm and 0.45 μm, respectively.

From these diffraction outputs, it can be observed that the central magnitude becomes larger as the aperture size expands. However, the magnitude around the edge region rises first and then decreases with the increase of the aperture size. A larger magnitude value on the edge region is favorable and a prominent magnitude gap between the central and edge regions is undesirable. Hence, the aperture with a size of 0.2 μm is chosen for the subsequent simulations.

Table 2. Optimization parameters of the D²NN

| Parameter | Value |
| --- | --- |
| Loss Function | MSE |
| Optimizer | SGD |
| Learning Rate | 0.01 |
| Momentum | 0.9 |
| Gamma | 0.7 |
| Batch Number | 64 |
| Epoch Number | 100 |

## 3.2 Optimization of the D$^2$NN

Various input plane waves with different tilt angles pass through the D$^2$NN, and the main target of this work is to constrain the output intensity to the small fixed central region, shown in Fig. 4. Therefore, it is a multiple-input single-output regression task. In the selection of the loss function, the mean square error (MSE) is used to measure the distance between the output and the target. During the backpropagation of model training, the stochastic gradient descent (SGD) algorithm is selected for gradient computing and parameters updating with a momentum of 0.9. Detailed optimization parameter settings are listed in Table 2.

The learning rate is initially set to 0.01 with a step decay rate (gamma) of 0.7 per 10 iterations. The batch number is 64 and the total iteration (epoch) number for the model optimization is 100. The rest of the optimization settings can be referred to [18]. In this work, all the simulations on the D$^2$NN are implemented by Python version 3.8.0 and PyTorch version 1.9.0, and finished on a desktop computer (AMD Ryzen 5 3600 CPU, NVIDIA GeForce RTX 3080 Ti GPU, and Ubuntu 20.04 LTS operating system).

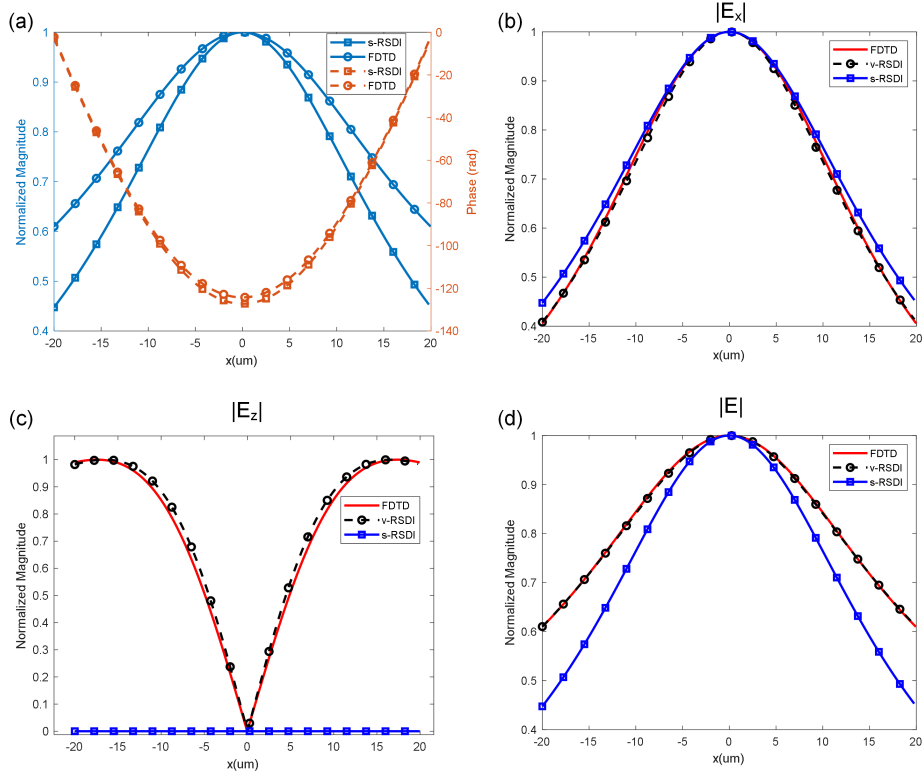

Fig. 6. Comparison of diffraction calculation results of an aperture with a size of 0.2 μm under the illumination at a wavelength of 0.45 μm among the FDTD, the v-RSDI, and the s-RSDI methods. (a) The normalized magnitude and phase along the x-coordinate of the FDTD and the s-RSDI methods. The normalized magnitude of (b) the component |E$_x$|, (c) the component |E$_z$|, and (d) both along the x-coordinate computed by the three methods.

## 4. Simulation results

### 4.1 Diffraction calculation

To verify that the scalar diffraction theory does not yield accurate diffraction calculations, a pair of contrast simulations are performed. Under the illumination at a wavelength of 0.45 μm, the diffraction output passing through an aperture with a size of 0.2 μm is calculated by the s-RSDI and the FDTD methods, respectively. The normalized magnitude and phase along the x-coordinate in the output plane are shown in Fig. 6 (a). Compared with the phase, there is a more salient gap in the magnitude distribution between the two methods, which is consistent with the previous theoretical inference that the s-RSDI is unsuitable to use when the aperture size is smaller than the light wavelength.

Under the same condition, several simulations are also performed to prove the correctness of the vectorial diffraction theory. The normalized magnitudes of different output electrical field components are shown in Fig. 6 (b)(c)(d). It is observed that the results of the v-RSDI method can almost fit the results of the FDTD method.

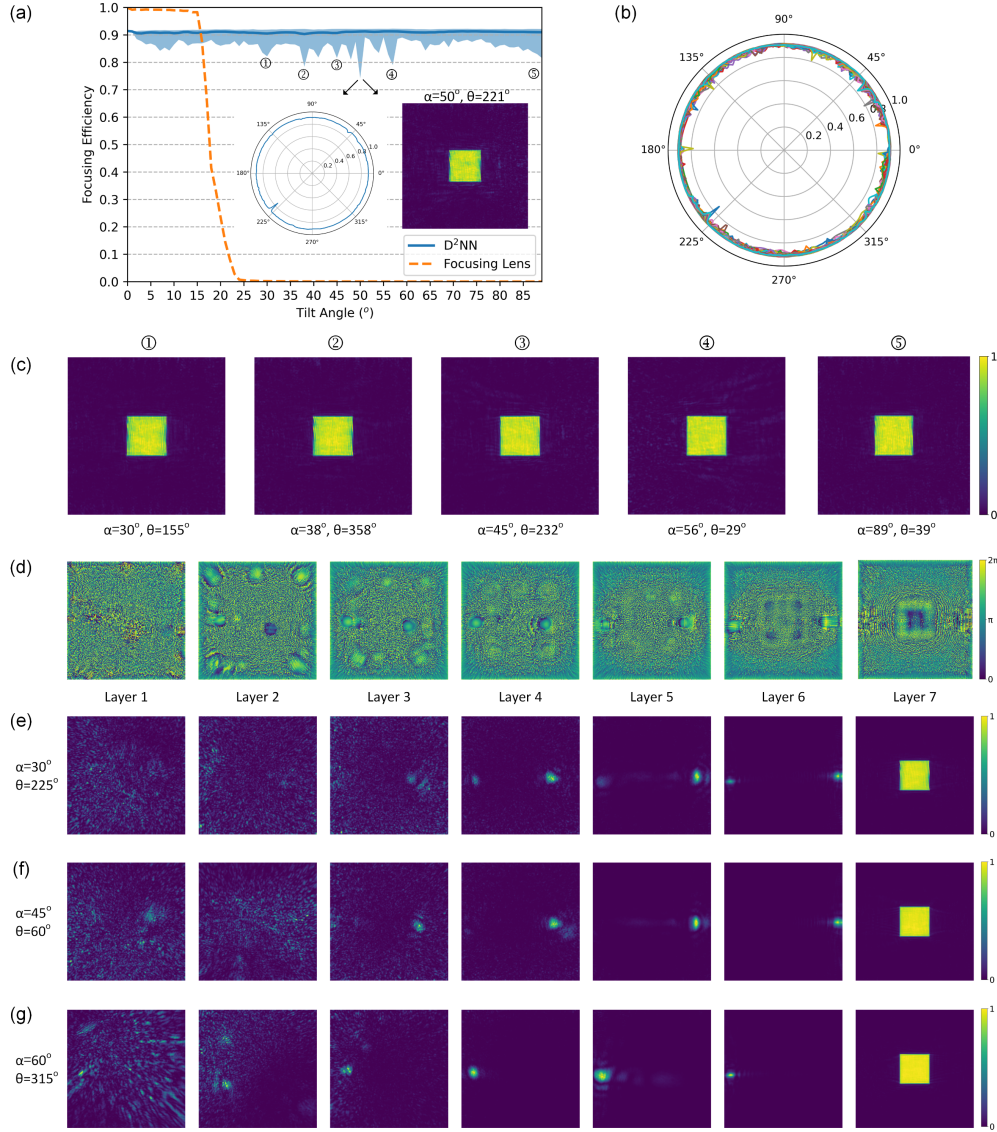

Fig. 7. Simulation results of omnidirectional focusing based on the $D^2NN$. (a) Comparison of focusing efficiency between the $D^2NN$ and a focusing lens. (b) The trajectory of focusing efficiency of all the outputs generated by the $D^2NN$ with different tilt angle $\alpha$ and rotation angle $\theta$. (c) Several output samples of the $D^2NN$ after normalization. (d) The optimized phase modulation parameters of each layer. Normalized intermediate intensity results in the processing procedure of the $D^2NN$ including (e) $\alpha=30°$ and $\theta=225°$, (f) $\alpha=45°$ and $\theta=60°$, (g) $\alpha=60°$ and $\theta=315°$.

## 4.2 Omnidirectional focusing based on the D²NN

In this work, the focusing efficiency is defined as the ratio of the intensity summation in the focusing region ($I_{region}$) to the total intensity on the output plane ($I_{total}$), which is written as

$$\tau = \frac{I_{region}}{I_{total}}. \tag{6}$$

After optimization, compared with a common focusing lens, although the focusing efficiency of the D²NN is slightly lower, the D²NN can achieve an average focusing efficiency of 90.96% among various incident waves with tilt angles from 0 to 89 degrees, as shown in Fig. 7(a). In a real submarine environment, the proposed method, which is highly integrated with detectors, can better achieve reliable link alignment in UWOC systems.

The average focusing efficiency curve (deep blue line) of the D²NN is smooth with some imperfections of performance variation (light blue region). Nevertheless, those flaws appear randomly at different angular positions and represent only a minority of the overall results, which is negligible and illustrated in Fig. 7(b). The imperfections mainly stem from the mapping constraints between the input and the output [32]. The current successive layer structure of the D²NN cannot provide enough degrees of freedom to control the point-to-point mapping from the input end to the output side.

For any tilt angle from 0 to 89 degrees, the D²NN can always restrict the output intensity to a small fixed central square region, as shown in Fig. 7(c). The optimized phase modulation parameters on different diffractive layers are listed in Fig. 7(d). It is difficult to figure out the function of each layer directly, but it can be inferred from the output intensities of each layer, as shown in Fig. 7(e)(f)(g).

For the incident waves with different tilt angles, the main function of layer 1 is to scatter the input light field. Then after passing through layer 2, 3, and 4, the intensity is always rearranged into the fixed region of the left or right side or both sides. And this kind of function is enhanced by the processing of layer 5 and 6. Finally, layer 7 converts the spot-like intensity into the target intensity, which is similar to the Fourier transform operation.

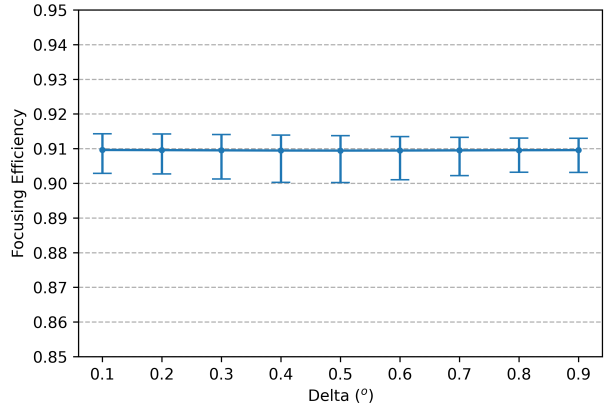

Fig. 8. Focusing efficiency results of incident waves with unused tilt angles for the 7-layer D²NN with a focusing region size of 50×50.

## 5. Discussion

### 5.1 Inference ability

In the above simulations, the resolution of the tilt angle is set to 1 degree and the omnidirectional focusing in this study is considered a regression task. Therefore, to verify that the optimized D²NN can work on continuous values of the tilt angle without over-fitting, a test is performed to let the optimized D²NN process incident waves with previously unused tilt angles. A higher resolution of the tilt angle is adopted and set to 0.1 degrees in the test. Over a range of 1 degree, nine values of the delta (Δ) are sampled with an interval of 0.1 degrees and added to the former tilt angles *α*, respectively, written as

$$\alpha' = \alpha + \Delta. \tag{7}$$

The incident waves with new tilt angles *α′* pass through the same D²NN and the focusing efficiency results are shown in Fig. 8. From the results, it shows that the mean focusing efficiency keeps stable at around 91% without much performance variation, which proves that the optimized D²NN can solve the omnidirectional focusing problem.

*5.2 Layer number*

In the structural design of the D²NN, it is hard to directly determine the effect of the layer number. Hence, multiple simulations with different layer numbers are performed and the results are shown in Fig. 9.

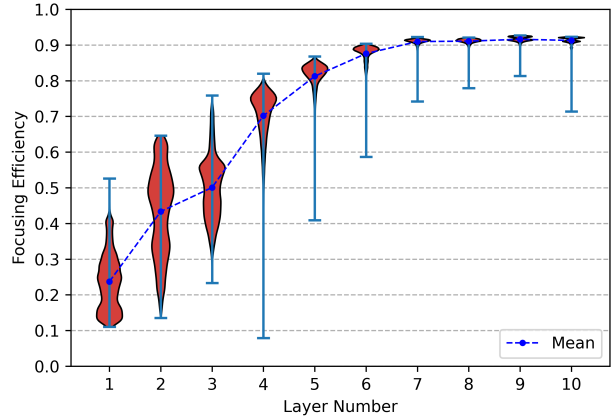

Fig. 9. The violin plot of the focusing efficiency of D²NNs with various diffractive layer numbers from 1 to 10 and a focusing region size of 50×50.

It can be observed that the focusing efficiency improves and becomes more centralized and compact with the increase of layer number. Meanwhile, the performance stabilizes at around 91% since the layer number reaches 7. The bottleneck of the focusing efficiency indicates that the performance enhancement carried by the increase of layer number is finite. As mentioned before, the limited degrees of freedom of the current layer structure should be the main cause.

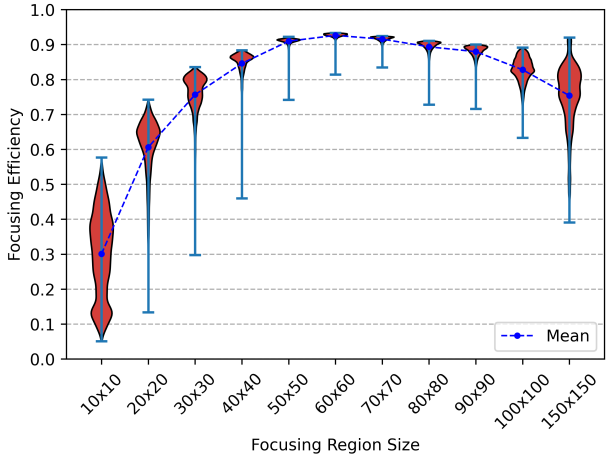

Fig. 10. The violin plot of the focusing efficiency for 7-layer D²NN with different focusing region sizes.

*5.3 Focusing region size*

As with the number of layers, the focusing region size is another parameter that is difficult to determine the effect. Therefore, simulations of different focusing regions are also performed and the results are illustrated in Fig. 10. The above results show that the mean focusing efficiency first rises and decreases later as the focusing region size increases. While the performance reaches the top when the focusing region size is 60×60, the average focusing efficiency maintains over 90% when the focusing region size is between 50×50 and 70×70. The results demonstrate that the $D^2NN$ with current structure works well in a certain range of focusing region size, which can be further improved by proper structural design and optimization.

## 6. Conclusion

In this paper, an omnidirectional receiver based on 7-layer $D^2NN$ is first proposed to alleviate the link alignment difficulties in UWOC systems, and the vectorial diffraction theory is introduced into the training of the $D^2NN$ to overcome the theoretical defect of the scalar diffraction theory. Compared with the widely adopted scalar diffraction theory, the vectorial diffraction theory can obtain more accurate diffraction calculations when the aperture size is smaller than the light wavelength. Moreover, the 9-point discrete sampling method is also proposed to simplify the diffraction computation. The simulation results demonstrate that the optimized $D^2NN$ can focus incident plane waves with tilt angles from 0 to 89 degrees in the focusing region at an average focusing efficiency of 90.96%. The performance of the $D^2NN$ peaks and keeps steady since the layer number reaches 7, and the 7-layer $D^2NN$ works well when the focusing region size is between 6.25% and 12.25% area of the detection plane. In the future, more research is needed to focus on large-scale fabrication of the $D^2NN$, structure simplification by improved optimization, complex underwater interference adaptation, etc. It is hopeful that both omnidirectional and reliable link establishment with high integration can be achieved in UWOC systems.

**Funding.** This work was supported by National Natural Science Foundation of China (NSFC) (61971378); Strategic Priority Research Program of the Chinese Academy of Sciences (XDA22030208); Zhoushan-Zhejiang University Joint Research Project (2019C81081).

**Disclosures.** The authors declare no conflicts of interest.

**Data availability.** Data underlying the results presented in this paper are not publicly available at this time but may be obtained from the authors upon reasonable request.

**References**

1. Z. Zeng, S. Fu, H. Zhang, Y. Dong, and J. Cheng, "A Survey of Underwater Optical Wireless Communications," IEEE Communications Surveys & Tutorials 19, 204-238 (2017).
2. H. Kaushal and G. Kaddoum, "Underwater Optical Wireless Communication," IEEE Access 4, 1518-1547 (2016).
3. J. Xu, "Underwater wireless optical communication: why, what, and how? [Invited]," Chinese Optics Letters 17, 100007 (2019).
4. X. Huang, F. Yang, and J. Song, "Hybrid LD and LED-based underwater optical communication: state-of-the-art, opportunities, challenges, and trends [Invited]," Chinese Optics Letters 17, 100002 (2019).
5. Z Tong, X Yang, X Chen, H Zhang, Y Zhang, H Zou, L Zhao, J Xu, "Quasi-omnidirectional transmitter for underwater wireless optical communication systems using a prismatic array of three high-power blue LED modules," Opt. Express 29, 20262-20274 (2021).
6. C. Yu, X. Chen, Z. Zhang, G. Song, J. Lin, and J. Xu, "Experimental verification of diffused laser beam-based optical wireless communication through air and water channels," Optics Communications 495, 127079 (2021).
7. X Li, Z Tong, W Lyu, X Chen, X Yang, Y Zhang, S Liu, Y Dai, Z Zhang, C. Guo, J Xu, "Underwater quasi-omnidirectional wireless optical communication based on perovskite quantum dots," Opt. Express 30, 1709-1722 (2022).
8. B. Han, W. Zhao, Y. Zheng, J. Meng, T. Wang, Y. Han, W. Wang, Y. Su, T. Duan, and X. Xie, "Experimental demonstration of quasi-omni-directional transmitter for underwater wireless optical communication based on blue LED array and freeform lens," Optics Communications 434, 184-190 (2019).

9. M. Zhao. X. Li, X. Chen, Z. Tong, W. Lyu, Z. Zhang and J. Xu, " Long-reach underwater wireless optical communication with relaxed link alignment enabled by optical combination and arrayed sensitive receivers," Opt. Express 28, 34450-34460 (2020).
10. Z Tong, X Yang, H Zhang, Y Dai, X Chen J Xu, "Series-connected solar array for high-speed underwater wireless optical links," Opt. Lett. 47, 1013-1016 (2022).
11. C. H. Kang, A. Trichili, O. Alkhazragi, H. Zhang, R. C. Subedi, Y. Guo, S. Mitra, C. Shen, I. S. Roqan, T. K. Ng, M.-S. Alouini, and B. S. Ooi, "Ultraviolet-to-blue color-converting scintillating-fibers photoreceiver for 375-nm laser-based underwater wireless optical communication," Opt. Express 27, 30450-30461 (2019).
12. M. Sait, A. Trichili, O. Alkhazragi, S. Alshaibaini, T. K. Ng, M.-S. Alouini, and B. S. Ooi, "Dual-wavelength luminescent fibers receiver for wide field-of-view, Gb/s underwater optical wireless communication," Opt. Express 29, 38014-38026 (2021).
13. Y. Guo, M. Kong, M. Sait, S. Marie, O. Alkhazragi, T. K. Ng, and B. S. Ooi, "Compact scintillating-fiber/450-nm-laser transceiver for full-duplex underwater wireless optical communication system under turbulence," Opt. Express 30, 53-69 (2022).
14. J Lin, Z Du, C Yu, W Ge, W Lyu, H Deng, C Zhang, X Chen, Z Zhang, J Xu, "Machine-vision-based acquisition, pointing, and tracking system for underwater wireless optical communications," Chinese Optics Letters 19, 050604(2021)
15. G. Wetzstein, A. Ozcan, S. Gigan, S. Fan, D. Englund, M. Soljacic, C. Denz, D. A. B. Miller, and D. Psaltis, "Inference in artificial intelligence with deep optics and photonics," Nature 588, 39-47 (2020).
16. C. Huang, V. J. Sorger, M. Miscuglio, M. Al-Qadasi, A. Mukherjee, L. Lampe, M. Nichols, A. N. Tait, T. Ferreira de Lima, B. A. Marquez, J. Wang, L. Chrostowski, M. P. Fok, D. Brunner, S. Fan, S. Shekhar, P. R. Prucnal, and B. J. Shastri, "Prospects and applications of photonic neural networks," Advances in Physics: X 7, 1981155 (2022).
17. N. H. Farhat, D. Psaltis, A. Prata, and E. Paek, "Optical implementation of the Hopfield model," Appl. Opt. 24, 1469-1475 (1985).
18. X. Lin, Y. Rivenson, N. T. Yardimci, M. Veli, Y. Luo, M. Jarrahi, and A. Ozcan, "All-optical machine learning using diffractive deep neural networks," Science 361, 1004-1008 (2018).
19. J. Xiong, Z. Zhang, and J. Xu, "Advances and progress of diffractive deep neural networks," in Applied Optics and Photonics China 2021, (SPIE, 2021), p. 120690V.
20. J. Li, D. Mengu, Y. Luo, Y. Rivenson, and A. Ozcan, "Class-specific differential detection in diffractive optical neural networks improves inference accuracy," Advanced Photonics 1, 046001 (2019).
21. T. Yan, J. Wu, T. Zhou, H. Xie, F. Xu, J. Fan, L. Fang, X. Lin, and Q. Dai, "Fourier-space Diffractive Deep Neural Network," Phys. Rev. Lett. 123, 023901 (2019).
22. Y. Sun, M. Dong, M. Yu, L. Lu, S. Liang, J. Xia, and L. Zhu, "Modeling and simulation of all-optical diffractive neural network based on nonlinear optical materials," Opt. Lett. 47, 126-129 (2022).
23. D. Mengu, Y. Zhao, N. T. Yardimci, Y. Rivenson, M. Jarrahi, and A. Ozcan, "Misalignment resilient diffractive optical networks," Nanophotonics 9, 4207-4219 (2020).
24. D. Mengu, Y. Rivenson, and A. Ozcan, "Scale-, Shift-, and Rotation-Invariant Diffractive Optical Networks," ACS Photonics 8, 324-334 (2021).
25. J. W. Goodman, Introduction to Fourier Optics (W. H. Freeman and Company, 2017).
26. R. K. Luneburg, Mathematical Theory of Optics (University of California Press, 1964).
27. A. Ciattoni, B. Crosignani, and P. Di Porto, "Vectorial analytical description of propagation of a highly nonparaxial beam," Optics Communications 202, 17-20 (2002).
28. N. Alcalá Ochoa, "A unifying approach for the vectorial Rayleigh–Sommerfeld diffraction integrals," Optics Communications 448, 104-110 (2019).
29. X. Luo, Y. Hu, X. Ou, X. Li, J. Lai, N. Liu, X. Cheng, A. Pan, and H. Duan, "Metasurface-enabled on-chip multiplexed diffractive neural networks in the visible," Light: Science & Applications 11, 158 (2022).
30. D. Mengu, Y. Zhao, A. Tabassum, M. Jarrahi, and A. Ozcan, "Diffractive interconnects: all-optical permutation operation using diffractive networks," Nanophotonics (2022). doi:10.1515/nanoph-2022-0358
31. F. Shen and A. Wang, "Fast-Fourier-transform based numerical integration method for the Rayleigh-Sommerfeld diffraction formula," Appl. Opt. 45, 1102-1110 (2006).
32. O. Kulce, D. Mengu, Y. Rivenson, and A. Ozcan, "All-optical synthesis of an arbitrary linear transformation using diffractive surfaces," Light: Science & Applications 10, 196 (2021).